\documentclass[epj]{svjour}
\usepackage{graphics}
\usepackage{amssymb}
\usepackage{epsfig}
\begin{document}
\title{Universal properties of shortest paths in isotropically correlated
random potentials}
\author{Roland Schorr and Heiko Rieger
}
\institute{Theoretische Physik, Universit\"at des Saarlandes,
66041 Saarbr\"{u}cken, Germany}
\date{Received: date / Revised version: date}
%
\abstract{
We consider the optimal paths in a $d$-dimensional lattice, where the
 bonds have {\it isotropically correlated} random weights. These paths
 can be interpreted as the ground state configuration of a simplified
 polymer model in a random potential. We study how the universal
 scaling exponents, the roughness and the energy fluctuation exponent,
 depend on the strength of the disorder correlations. Our numerical
 results using Dijkstra's algorithm to determine the optimal path in
 directed as well as undirected lattices indicate that the
 correlations become relevant if they decay with distance slower than
 $1/r$ in $d=2$ and $3$. We show that the exponent relation
 $2\,\nu-\omega=1$ holds at least in $d=2$ even in case of
 correlations. Both in two and three dimensions, overhangs turn out to
 be irrelevant even in the presence of strong disorder correlations.
} 
\maketitle

\section{Introduction}
Optimal paths have been a subject of intensive studies during the
recent years. Besides being one of the simplest problems involving
disorder, this interest can be traced back to the relevance of this
problem to various fields, such as polymer models
\cite{ref10,ref16,ref26}, surface growth 
\cite{ref6}, random bond ferromagnets
\cite{ref2,ref3,ref4}, spin glasses \cite{ref1}, and the traveling
salesman problem \cite{ref5}. 

The model under consideration is easily sketched: given an arbitrary
weighted graph, each edge has a particular cost. The optimal
or shortest path connecting two sites is the one of minimal
weight, which is the sum of all costs along that path. We do not
restrict to a particular geometry yet as well as we do not specify the
costs more precisely so far. In the simplest case, we choose them from
a set of random numbers that are uniformly distributed. In this
context, the {\it directed} polymer model has drawn the most
significant attention during the past years
\cite{ref10,ref16,ref7,ref8,ref9}, where one assumes in $d=2$ a simple
square lattice being cut along its diagonal and oriented as a triangle
with the diagonal as its base. One allows only paths in the
direction to the base, i.e., the path cannot turn backwards. The costs
of the edges that belong to the shortest path are interpreted as
potential energies for a polymer configuration that passes through
these edges (or bonds).

We may now ask whether the properties (scaling exponents) of the
shortest path are either influenced by the distribution of the random
numbers or the geometry of the lattice. The former is still
discussed \cite{ref10,ref11,ref12,ref13,ref14}. 
As far as the latter is concerned, it seems to be clear that the
universal properties are not changed if the randomness is uncorrelated. 
For this case Schwartz \emph{et al.} \cite{ref15}
investigated directed and undirected paths in $d=2,3$
using Dijkstra's algorithm to find the shortest path and Marsili
and Zhang \cite{ref27} used a transfer matrix method approach
considering directed and undirected paths up to $d=6$. Both state that
overhangs exist but nevertheless they suggest that both problems belong
to the same universality class,
even in high dimensions where overhanging configurations are
entropically favored. It is not a priori clear that this observation
remains true for correlated disorder. In fact, as we will point out
below, the average number of overhangs increases for strongly correlated
disorder indicating that they might become relevant for strong enough
correlations.

In the present study, we study the universal properties of shortest
paths and focus on the effect of isotropically correlated random
weights on the scaling exponents. To this end, we consider directed
and undirected lattice-graphs in two and three dimensions with bond
weights $\eta_{\bf j}$, where the $d$-dimensional index vector ${\bf
j}={i_1,...,i_d}\in{\mathbb Z}^d$ denotes the position of a particular
bond in the lattice. The total energy or cost of a path ${\cal P}$
from one end of the lattice (e.g.\ from one special site or node in the
top layer) to the opposite end (e.g.\ to an arbitrary site or node in
the bottom layer) is simply the sum of these bond weights
\begin{equation}\label{eq1}
E=\sum_{{\bf j}\in{\cal P}} \eta_{\bf j}\;.
\end{equation}
The weights $\eta_{\bf j}$ are {\it correlated} positive random
variables, which we define below. We choose the index vector ${\bf j}$
of the bonds in such a way that it is identical with the position of
the center of the bonds in an Euclidean lattice. E.g. for the square 
lattice ($2d$)
in which we define the upper left corner as the origin, all indices
take on odd values. We refer to isotropic correlations if the
connected correlation function of the costs $\eta_{\bf j}$ decays with
a power law with Euclidean distance of two bonds, viz.
\begin{equation}
G({\bf j},{\bf j}')=\langle\eta_{\bf j}\eta_{{\bf j}'}\rangle-\langle
\eta_{\bf j}
\rangle\langle\eta_{{\bf j}'}\rangle \sim |\,{\bf j}-{\bf
  j}'\,|^{2\rho-1},
\label{eq2}
\end{equation}
where $\rho<1/2$. Here $\langle\cdots\rangle$ denotes the disorder
average, i.e. an average of the probability distribution of $\eta_{\bf j}$.

For $\rho=0$ model (\ref{eq1}) with (\ref{eq2}) represents an
effective single vortex line model for interacting vortex lines in a
random vector potential \cite{natter}. We would like to emphasize that
{\it isotropic} correlations have not been investigated so far: For
historical reasons one discriminates between $d-1$ transverse or
spatial directions and 1 longitudinal or time direction. This can be
traced back to the relation between directed polymers and the
KPZ-equation. If the randomness is only correlated in $d-1$ dimensions
and uncorrelated in the remaining direction one refers to {\it spatially}
correlated randomness, if the correlations are only present in time
direction one refers to {\it temporal} correlations. Let ${\bf j}=({\bf
x},t)=((i_1,...,i_{d-1}),t)$.  Then we can describe both cases by
$G\sim|\,{\bf x}-{\bf x}'\,|^{2\rho-1}\cdot\delta(t-t')$ (spatial
correlations) and $G\sim|\,t-t'\,|^{2\theta-1}\cdot\delta({\bf x}-{\bf
x}')$ (temporal correlations).

The first systematic numerical work done on spatially
correlated noise is due to Amar, Lam and Family \cite{ref22}. Their
results are in agreement with the dynamical \cite{ref23} and functional RG
\cite{ref24,ref25} predictions. Subsequent careful work by Peng 
\emph{et al.} \cite{ref21} and Pang \emph{et al.} \cite{ref17} was done.
As far as temporal correlations are concerned, we refer to the
simulation of a ballistic deposition model by Lam, Sander and Wolf
\cite{ref20}.

In this paper we study {\it isotropically} correlated disorder, it is
organized as follows: In section II we introduce the models, the
numerical method and the quantities that we are interested in. In
section III we present our results and in section IV we summarize our
findings.

One remark on the notation: We will use the words graph and lattice as
well as costs and energy, node and site and edge and bond synonymously
throughout the paper.

\section{Model\label{sectwo}}
\begin{figure}[hbt]
\centerline{\epsfig{file=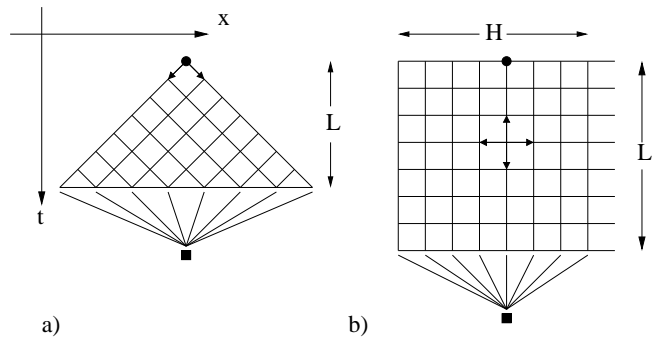,scale=0.6}}
\caption{a) The {\it directed} graph can be regarded as a
 square lattice that was cut along its diagonal and oriented as a
 triangle with the diagonal is its base. The paths are directed
 downwards to the base. b) In the {\it undirected} case the path is
 allowed to turn back on a lattice with periodic boundary conditions
 in the spatial direction(s). In both cases one fixes the source
 {\Large$\bullet$}, whereas the target $\blacksquare$ is the most
 favorable node of the base.}
\label{nlat} 
\end{figure}

The \emph{undirected graph} can be described as follows (see Fig.\ref{nlat}b)):
we choose a simple lattice structure and define one longitudinal and
$d-1$ transversal directions. We will refer to them by means of $t$ and
${\mathbf x}$ respectively. We assume the lattice to be periodic in
space (${\mathbf x}$) with period $H$ and $L$ to be the longitudinal
size. We choose the origin of the coordinate system as being the starting
point (the source) of the path. We assign a particular amount of energy
to each bond, whereby these
energies are isotropically correlated. Generating these random numbers
we follow the method introduced by Pang \emph{et al.} \cite{ref17}. 
We infer periodicity and symmetry of the correlator $g_\rho$ in any
direction, where 
\begin{equation}\label{corln}
g_\rho(\Delta {\bf j}):=G(\,{\bf j},{\bf j}+\Delta{\bf j}\,)=
|\Delta {\bf j}|^{2\rho-1}
\end{equation}
with $g_\rho(...,(\Delta {\bf j})_i+p_i,...):=g_\rho(...,(\Delta {\bf
j})_i,...)$ and \\ $g_\rho(...,(\Delta {\bf
j})_i,...)=g_\rho\,(...,p_i/2-|\,(\Delta {\bf j})_i-p_i/2\,|,...)$ in
case of a period $p_i$ in direction $i$. 
In contrast to pure spatial and temporal correlations where the
correlator is taken to be the product of two separate ones, one for the
time direction and one for the remaining spatial coordinates, here
it is due to a generic vector $\Delta{\bf j}$. Adapting the correlator to
the lattice, we require a period $H$ in the transversal and $2L$ in
the longitudinal direction. By means of the factor $2$ we guarantee
that $g_\rho(t)$ has the required form (\ref{corln}) in the range 
$1\le t\le L$. Its Fourier
transform yields $S_\rho({\bf k})$ such that the correlator in the
${\bf k}$-space is given by $\langle\eta_{\bf k}\eta_{\bf
k'}\rangle\sim\delta_{{\bf k}+{\bf k'},0}\,S_\rho({\bf k})$. Choosing
$\eta_{\bf k}\equiv\sqrt{S_\rho({\bf k})}(r_{\bf k}-1/2)\,\exp(2\pi
i\phi_{\bf k})$, that relation can be fulfilled, where $r_{\bf k}$ and
$\phi_{\bf k}$ are random variables uniformly distributed between $0$
and $1$. A transformation back to real space provides the random
numbers correlated according to the power law rule (see also Appendix
\ref{aclat}).

From the set of $L^{d-1}$ optimal paths that connect the source $({\bf
x},t)=({\bf 0},0)$ with nodes of the bottom layer with coordinates
$({\bf x},L)$, we select the shortest one. Technically this is achieved
not by repeating the same calculation $L^{d-1}$ times (i.e.\ once for
each end-point) but by introducing an extra target node connected to
the bottom layer by zero-weight bonds. The algorithm that computes the
optimal paths in polynomial time is Dijkstra's algorithm that works in
any graph with non-negative weights.

Our study is focused on the universal characteristics
of the optimal path, i.e. on the scaling exponents $\nu$, the roughness
exponent, and $\omega$, the energy fluctuation exponent. $\nu$
describes the fluctuations of the path with
regard to a line parallel to the $t$-axes that is shifted to the
origin by an amount matching to the mean position $\langle{\bf
x}\rangle$ of the polymer.  We refer to these fluctuations by $D$. Due
to the direct mapping between the model of shortest paths and growing
interfaces \cite{ref6}, it is immediately seen that the energy of the
polymer also fluctuates, where we consider several realizations of
disorder: we have
\begin{equation}\label{scaling}
D\sim t^\nu\qquad\qquad\Delta E\sim t^\omega
\end{equation}
There is no need to modify these relations when we consider
\emph{directed paths}. For this purpose we introduce directed bonds, e.g., we
restrict to the positive axes. In order to determine the roughness of the
polymer, we refer to a line parallel to the bisecting line $({\bf
x},t)\sim(1^{d-1},1)$ that crosses the origin. For each size $L$ we can
determine the distance of the target node to that line by considering
the projection of its position vector onto the bisector. That
is, a directed path connects the source $({\bf x},t)=({\bf 0},0)$ and
the sites on the line between $(0,L)$ and $(L,0)$, what has to be
extended to the notion of a plane in $d=3$. 

As far as undirected paths are concerned one has to make clear how 
${\bf x}(t)$ can be defined if overhangs appear. In that case 
${\bf x}(t)$ is not a single valued quantity anymore. We have checked
the relevance of several choices but finally we took ${\bf
x}(t)=\mbox{max}\{{\bf x}_i(t)\}$ with $t$ constant for all ${\bf
x}_i$. The numerical results that we present are independent of this choice.

As there are ${\mathcal O}(2^n)$ undirected paths across the lattice,
where $n$ is the number of 
nodes, an efficient method is needed that terminates within a reasonable
time, such as Dijkstra's algorithm \cite{ref15,ref18}, that is able to
generate the shortest path in polynomial time. In this algorithm the path is
successively constructed, i.e., one obtains not only a single path but a
cluster of them with energy labels smaller than a certain limit, so that
a growth front is established. This cluster contains the so called
permanently labeled sites. The algorithm proceeds by extending the front
by this site that is the nearest neighbor to it with respect to the
energy. If the growth front reaches the base of the lattice we are
enabled to reconstruct the shortest path. The growth process of that
front can directly be mapped onto a growing Eden cluster \cite{ref6}
providing the same scaling behavior as directed polymers.

If we assume uncorrelated costs the roughness of the growth
front scales like the energy fluctuations of the polymer. Since the
properties of the surface are described by the KPZ-equation and,
consequently, the height-height correlation increases according to a
power law with exponent $\omega=1/3$, we expect in $d=2$ the scaling
exponents of the optimal path of size:
\[ \omega_{\mbox{\tiny OP}}=\beta_{\mbox{\tiny
KPZ}}=1/3\qquad\nu_{\mbox{\tiny OP}}=1/z_{\mbox{\tiny KPZ}}=2/3 \]
Exact values are only accessible in $d=2$. The value of the
exponent $\nu$ can be extracted from $\omega$ if we take into account
the exponent relation \[2\,\nu-\omega=1\] which holds for uncorrelated
noise. This relation follows from the Galilean invariance of the
KPZ-equation and the transfer to shortest paths afterwards. In $d=3$
there exist no analytical predictions. Nevertheless, the estimates of
numerous numerical studies \cite{ref16,ref10,ref15} yield
\[ \omega\approx 0.19\qquad\qquad\nu\approx 0.62 \]
As long as the Galilean invariance holds, the scaling relation remains
unchanged. This invariance is not altered in the presence of spatial
(transversal) correlations but it is broken in the presence of temporal
(longitudinal) correlations.

A Flory type argument \cite{ref8} leads to the following estimate $\nu_F$
of the roughness exponent as a function of $\rho$. A continuum
Hamiltonian of the energy (\ref{eq1}) has to include an elastic part,
since this is generated in a coarse graining procedure.
\begin{equation}
{\mathcal H}=\int dt\,\left[\frac{\lambda}{2}(\nabla_t\,{\bf x})^2
+\eta({\bf x},t)
\right].
\end{equation}
Rescaling $t$ with a factor $b$ and $x$ with a factor $b^\nu$
the elastic term scales with $b\,^{2\nu-2}$ and the
disorder term with $b\,^{\rho-1/2}$ (because of the power law
decay of the disorder correlations), resulting in
\begin{equation}
\nu_F=\frac{1}{2}\,\rho+\frac{3}{4}.
\end{equation}
In case of uncorrelated random numbers the disorder term scales
with $b\,^{-\nu(d-1)/2-1/2}$, resulting in
\begin{equation}
\nu_{uncorr.}=\frac{3}{d+3}.
\end{equation}
Hence we expect
\begin{equation}
\nu=\left\{
\begin{array}{lcl}
\nu_{uncorr.} & \qquad{\rm for}\qquad & \rho\le\frac{3(1-d)}{2(3+d)}\\
\nu_F         & \qquad{\rm for}\qquad & \rho>\frac{3(1-d)}{2(3+d)}
\end{array}
\right.
\label{flory}
\end{equation}
This simplified scaling picture should yield at least a lower bound for
the roughness exponent.

\section{Numerical results}

In addition to the relations in (\ref{scaling}), we define the two exponents
$\gamma$ and $\delta$ by $l\sim L^\gamma$ and $B\sim L^\delta$, where
$B$ is the number of backward bonds with respect to time, and $l$ is
the total length of the path. We also determine the fractal dimension
$d_c$ of the shortest path cluster $M\sim L^{d_c}$, where $M$ is the
mass of its surface.  The shortest path cluster consists of all nodes
with labels smaller than a maximal one given by the shortest path
weight from the source to the base.  As far as Dijkstra's algorithm is
concerned, its surface is constituted by all the sites that are part
of the growth front with at least one nearest neighbor that is not yet
permanently labeled.

\begin{figure}[hbt]
\centerline{\epsfig{file=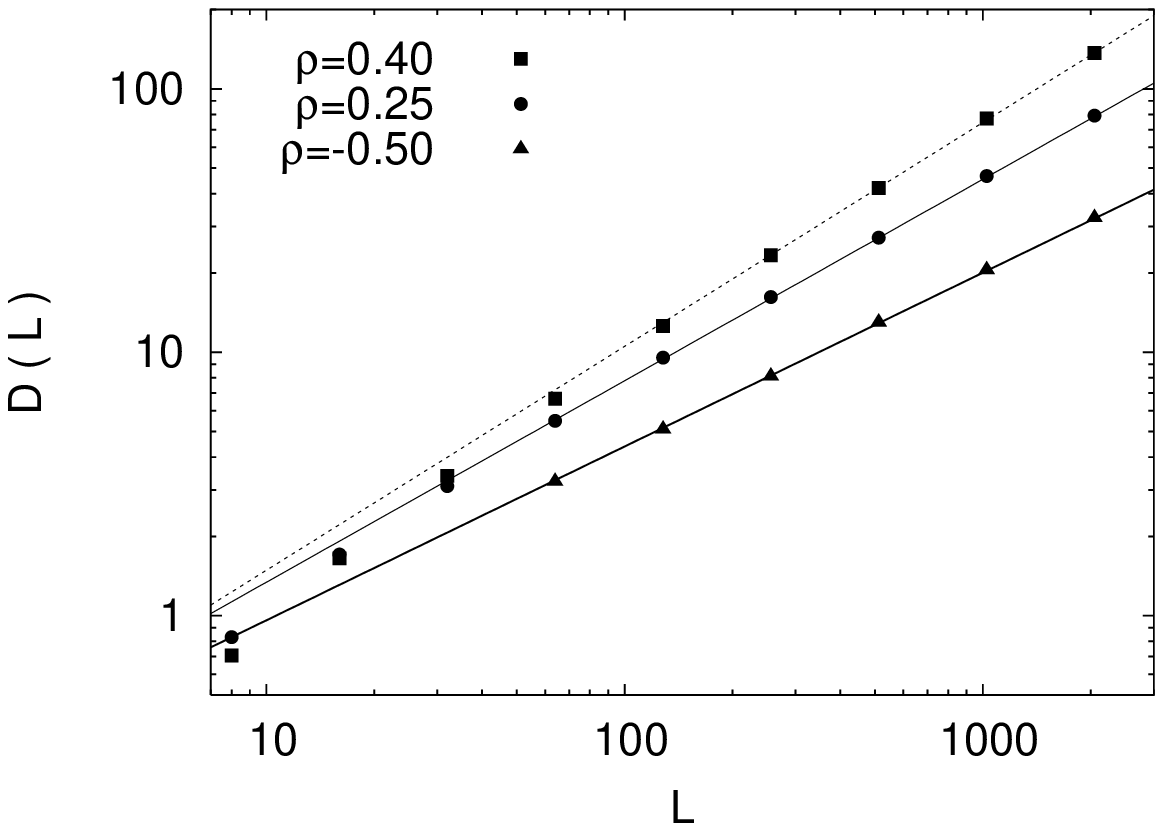,scale=0.69}}
\centerline{\epsfig{file=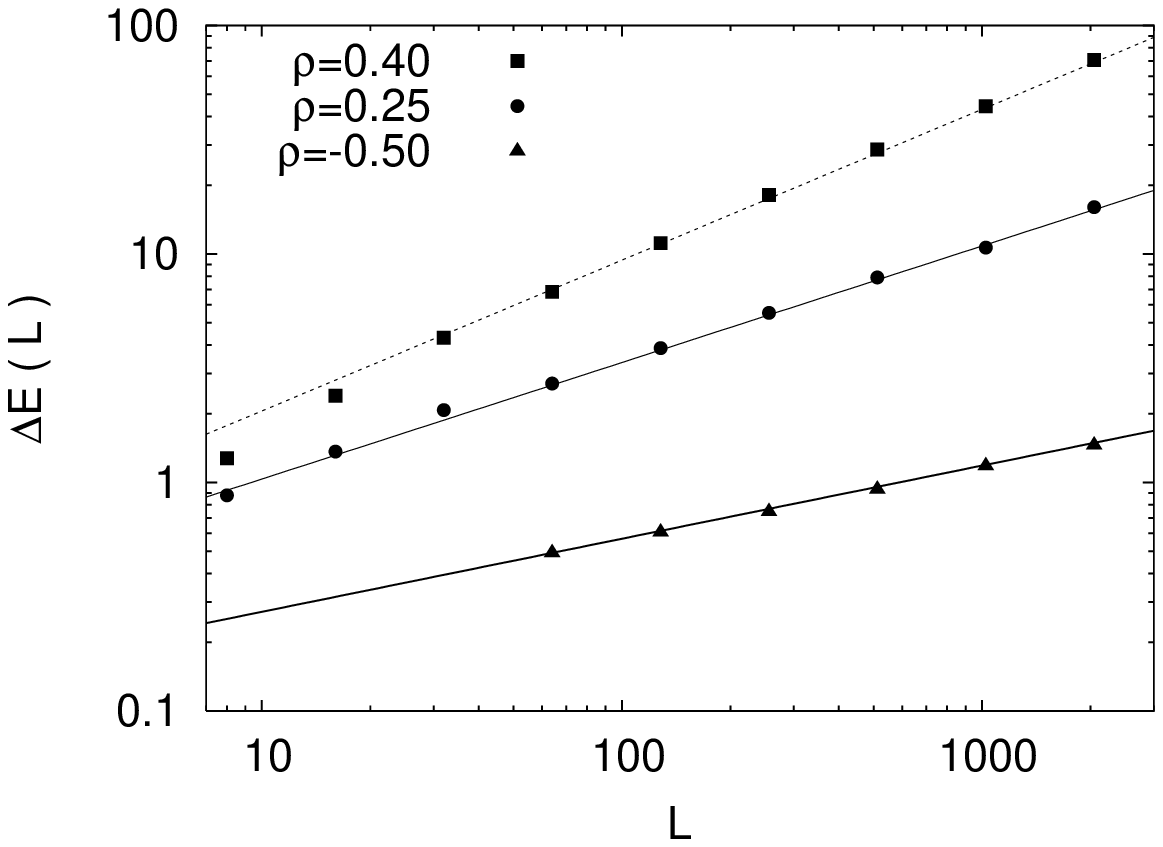,scale=0.69}}
\caption{Scaling of the height-height correlation $D(L)$
 and the energy fluctuations $\Delta E(L)$ in $d=2$ for NDOPs. The
 straight lines are least square fits of the data for $L>100$ to a power
 law $D(L)\propto L^\nu$ and $\Delta E(L)\propto L^\omega$ with
 $\nu=0.66,0.77,0.85$ and $\omega=0.32,0.50,0.67$ for
 $\rho=-0.50,0.25,0.40$. The statistical error of the data for $D(L)$
 and $\Delta E(L)$, which are averaged over at least 10000 disorder
 realizations, is smaller than the symbol size.}
\label{scales} 
\end{figure}

The scaling exponents are extracted from a set of data that reproduces
the simulation of several lattice sizes (Fig.\ref{scales}). 
The statistical error is usually smaller than the symbol size. We adapt
the transversal expansion $H$ to the size $L$ in such a way that we
eliminate further effects on our data by increasing $H$, even if
$\rho=\rho_{max}$. Finally, we consider lattices of size $H\ge 4L$ if
$L\le2048$ and $H=4096$ if $L=2048$ in $d=2$, as well as $H=128$ if
$L\le32$ and $H=256$ if $L\ge 64$ in $d=3$. At any time, we forbid the
shortest path cluster spanning the lattice in order to avoid
saturation effects. This requirement cannot be satisfied for all $L=128$
in case of $d=3$ and $\rho$ close to $\rho_{max}$.

Our results are averages over more than $10000$
disorder realizations per size $L$. The generation of
correlated random numbers is the most time consuming part. For each sample
we have to perform a Fourier transformation of $N=2^d LH^{d-1}$ numbers
twice. On the average approximately $90\mbox{\%}$ of the CPU time is
needed for this execution. Some more information about the generation is
given in Appendix \ref{aclat}.

\begin{figure}[hbt]
\centerline{\epsfig{file=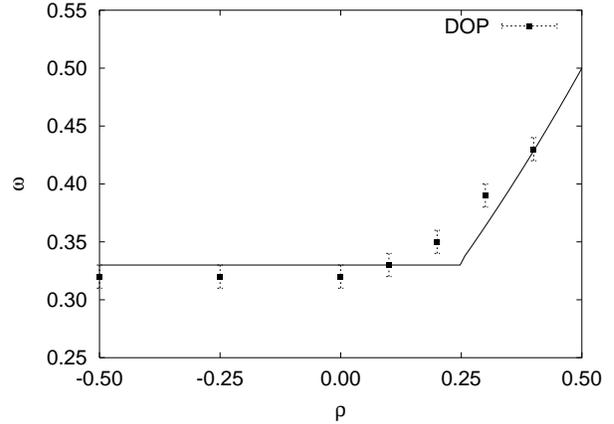,scale=0.65}}
\caption{\label{space} Our result of the directed path in $d=2$
 for spatial correlations. A direct comparison with the results of 
 \protect{\cite{ref17}} (Fig.3) shows that they are in good agreement.
 The full line is theoretical prediction from a
 one-loop KPZ dynamic RG calculation \protect{\cite{ref23}} and a DPRM
 functional RG \protect{\cite{ref25}} calculation.}
\end{figure}

As a first check we studied the directed polymer problem in $d=2$ 
with only spatially correlated bond weights as it was done by Pang
\emph{et al.} \cite{ref17}. It can be seen from Fig.\ref{space} that we
are in very good agreement with their results. In case of strongest
correlations ($\rho=0.40$) we obtain $\nu=0.71\pm0.01$ and
$\omega=0.43\pm0.01$. It is quite evident that correlations are only
relevant in the regime $\rho>0$.

\subsection*{Two dimension ($d=2$)}

\begin{figure}[hbt]
\centerline{\epsfig{file=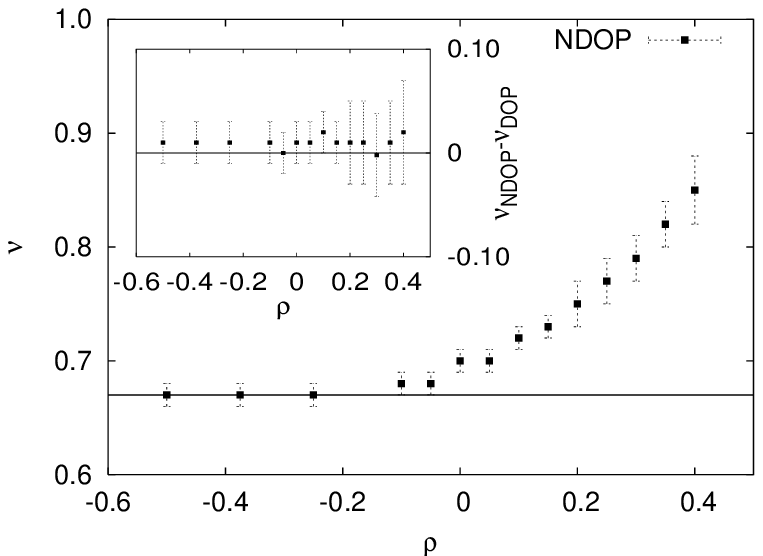,scale=1.03}}
\centerline{\epsfig{file=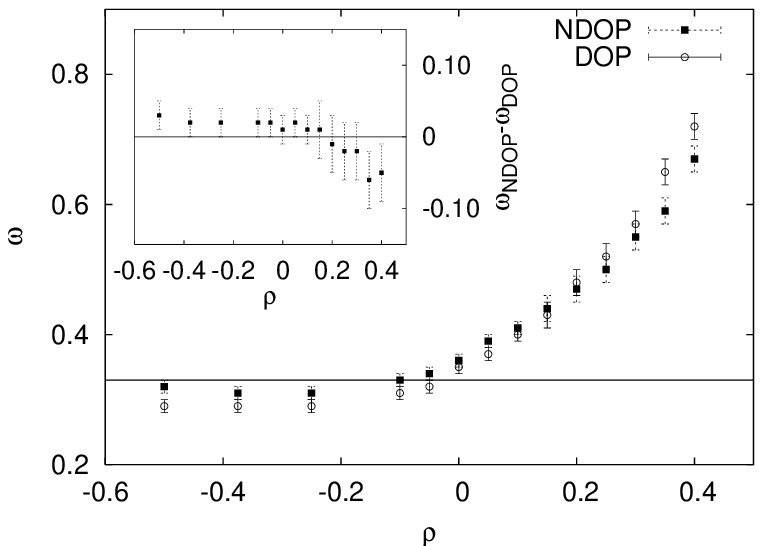,scale=1.03}}
\caption{\label{nyom2} The scaling exponents $\nu$ and
 $\omega$ in $d=2$ depend on the strength $\rho$ of the
 correlation. Close to $\rho=0$ correlations start to affect them
 significantly. In either case the insets show the difference of both
 the exponents of the undirected and directed lattice respectively. The
 reference lines represent the values $\nu=1/3$ and $\omega=2/3$
 respectively.}
\end{figure}

In $d=2$ we find that the roughness exponent $\nu$ does not depend
on the directedness of the lattice. Overhangs do not play an important role
and the undirected path can be regarded as a directed one
(Fig.\ref{nyom2}). Independent from the strength of the correlations all
exponents $\nu$ are smaller than the critical value 1. As far as
undirected shortest paths are concerned, exceeding that critical value
leads to fractal objects that cannot become directed, even on large
scales. The errorbars 
depicted in Fig.\ref{nyom2} are not the result of the least square fit
but estimates of the minimal and maximal slopes being in nearly perfect
agreement with our data.

\begin{figure}[hbt]
\centerline{\epsfig{file=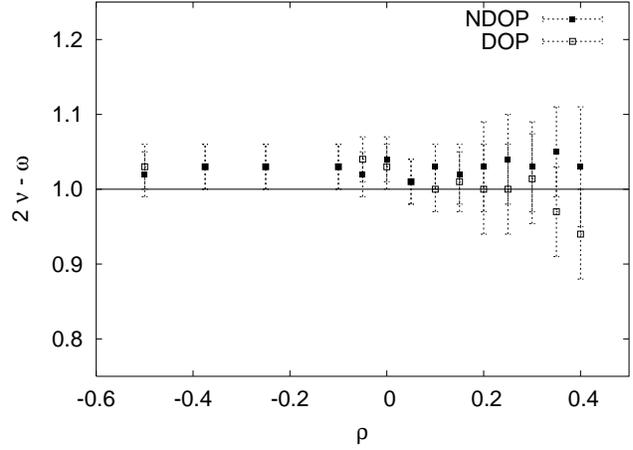,scale=0.7}}
\caption{\label{ident} The exponent relation is depicted for the NDOP
 model in $d=2$. Both data sets diverge only close to $\rho=0.40$
 according to the insets of Fig.\ref{nyom2}.}
\end{figure}

In contrast, we obtain a less significant data collapse with respect to
the energy fluctuation exponent
$\omega$, if $\rho>0$. This may be affected by a statistics that has room
of improvement but it indicates a tendency that is especially noticeable
in $d=3$: the stronger the correlation the more significant variations
in $\omega$ occur. The reason for this relation is not clear to us.
In both cases the exponent relation can be satisfied
(Fig.\ref{ident}) where we are in better agreement with
$2\,\nu-\omega=1$ for undirected paths. We learn from Fig.3 that
isotropic correlations in the randomness are relevant for $\rho>0$. The
scaling of the energy 
fluctuations is much more sensitive to passing from white noise to weak
correlations ($\rho=-0.5$) in comparison to the roughness. Whereas $\nu$
keeps constant ($\nu=0.66\pm0.01$), $\omega$ reduces from
$\omega=0.32\pm0.01$ to $\omega=0.29\pm0.01$. The number of backward
bonds in the NDOP model remains negligibly small. It is $B\approx15$ for
$L=1024$ if $\rho=0.40$ where $\delta\approx1.0$. In that case almost
every path has at least one such bond. According to these results we
obtain $l\sim L$.

\begin{figure}[hbt]
\centerline{\epsfig{file=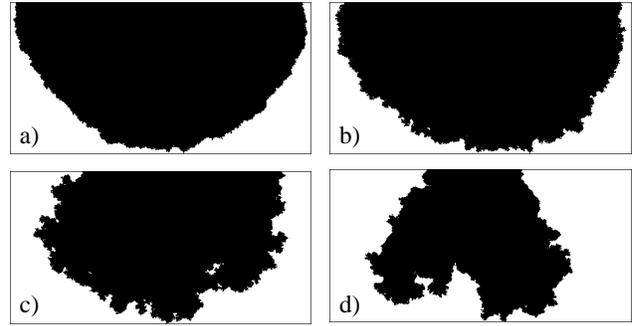,scale=1.25}}
\caption{\label{cl2} Realizations of the shortest path cluster
(NDOP) for $H=512$ and $L=256$. In each case the growth
 process stops when the growth front reaches the base of the lattice. We
 choose a) uncorrelated random numbers and correlated random numbers
 with b) $\rho=0$, c)+d) $\rho=0.40$.}
\end{figure}

Although the roughness exponent increases significantly for $\rho\ge0.3$
and might even come closer to 1 for $\rho\approx0.6$ it stays still
smaller than one. Also the length of the shortest path scales linearly
with $L$ for the values of $\rho$ that we could study. Both observations
imply a non-fractal shortest path. However, the properties of the
shortest path cluster change remarkably when coming closer and closer to
$\rho=0.40$, where the fractal dimension of its surface becomes $d_c\approx1.1$
instead of $d_c\approx1$ if $\rho=0$. In Fig.\ref{cl2} one can see that
for weak correlations the surface of the shortest path cluster is a
semicircle whereas for larger correlations it becomes topologically more
complicated.

\begin{figure}[hbt]
\centerline{\epsfig{file=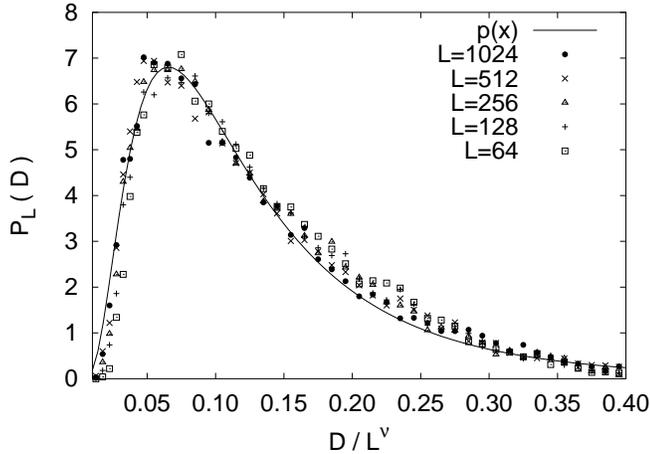,scale=0.7}}
\caption{\label{prob} Scaling plot for the
 probability distribution $P_L(D)$ of the roughness in case of directed
 paths for $\rho=0.40$. The scaling function $p(x)$ is a log-normal
 distribution with $D_0=0.11$ and $\sigma=0.70$.}
\end{figure}

Not only the disorder averaged roughness $D$ scales with $L^\nu$ but the
whole probability distribution $P_L(D)$: It is $P_L(D)=L^\nu p(D/L^\nu)$
as we show in Fig.\ref{prob} for $\rho=0.40$. For the scaling function
$p(x)$ we can fit a log-normal distribution given by
$p(x)=(2\pi\sigma^2)^{-1/2}\exp(-(\ln D/D_0)^2/2\sigma^2)$. 
The parameters $\sigma$ and $D_0$ do not depend on $\rho$ for
$\rho\le0$ (where $D_0\approx 0.13$, $\sigma\approx 0.58$) and vary
slightly with $\rho$ for $\rho>0$.

\subsection*{Three dimensions ($d=3$)}
\begin{figure}[hbt]
\centerline{\epsfig{file=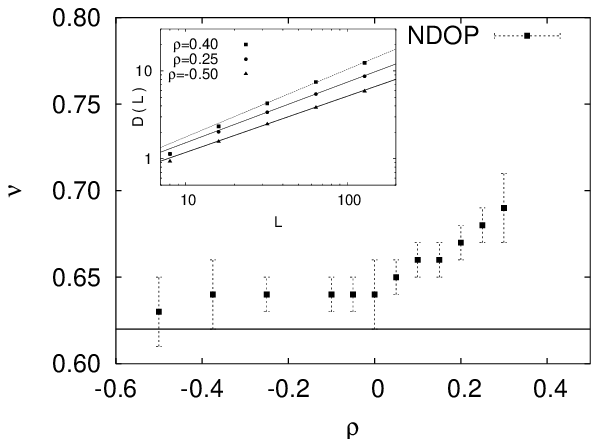,scale=1.42}}
\centerline{\epsfig{file=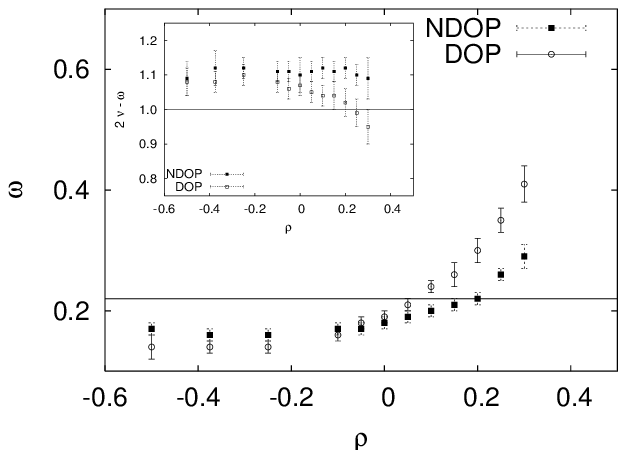,scale=1.35}}
\caption{\label{nyom3} The exponents $\nu$ and $\omega$ in $d=3$ are
 plotted versus 
 $\rho$. The insets show the scaling of the height-height
 correlation $D(L)$ for NDOP and the exponent relation in $d=3$
 respectively. The reference lines belong to our 
 results $\nu=0.62$ and $\omega=0.22$ in case of uncorrelated numbers.}
\end{figure}

The results in $d=3$ are qualitatively similar to those of the
preceeding section. For uncorrelated randomness we obtain $\nu=0.62 \pm
0.02$ and $\omega=0.22 \pm 0.01$ in agreement with
\cite{ref10,ref16,ref15}. Both $\nu$ and $\omega$ increase monotonuously
with $\rho$, i.e.\ increasing correlations. While $\nu$ does so without
any difference between directed and undirected paths, $\omega$ differs
from this behavior: the stronger the correlations the more estimates for
the exponents deviate from each other (Fig.\ref{nyom3}).
As $\nu<1$ the undirected path becomes directed on large scales. We
should emphasize that obtaining data in the regime $\rho>0.25$ for $d=3$
is much more delicate than in $d=2$. The local slopes of the energy
fluctuations indicate that even for 
the maximal size $L=128$ we are not yet in the asymptotic regime for
$\rho>0.3$. Therefore we explicitly restrict ourselves to values
$\rho\le0.3$. This result refers to both kinds of lattices and, so is not a
consequence of saturation effects. The distinct behavior of both kinds
of paths can be demonstrated more obviously by plotting the exponent
relation $2\nu-\omega=1$ (see inset of Fig.\ref{nyom3}). 

Even though the scaling regimes in $d=3$ are not very wide our estimates
for $\nu$ and $\omega$ deviate significantly from their values in the
uncorrelated case. Significantly means here that their estimated errors
(obtained in the same way as in our $2d$-study before) are smaller than
the deviation from the uncorrelated values. Therefore we can infer that
for strong enough correlations ($\rho\ge0.3$) the DOP and NDOP problems
constitute new universality classes. The precise determination of the
critical value for $\rho$, beyond which correlations become relevant,
is, however, beyond our numerical precision.

For completeness we mention that even in case of $L=128$ and
$\rho=0.40$ the number of bonds turning backward is negligibly
small: $B\approx1.6$ compared to $B\approx3$ in $d=2$. We do not expect a
significant change in $B$ for larger lattices as this quantity scales by an
exponent $\delta\approx1.0$ but the number of paths including such bonds
should increase from $60\%$ to $100\%$. 

\begin{figure}[hbt]
\centerline{\epsfig{file=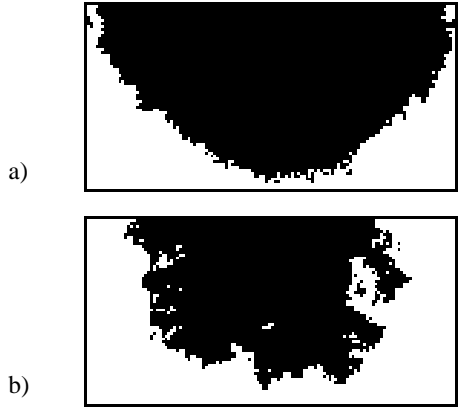,scale=1.68}}
\caption{\label{cl3} Cluster of shortest paths in $d=3$ that is cut
 along the $x-t-$ plane with $y=H/2$. a) uncorrelated randomness b)
 strong correlations ($\rho=0.40$). It is seen that we pass again from a
 semicircle like cluster to a topologically more complicated one.} 
\end{figure}

In contrast to the difficulties above, the fractal dimension $d_c$ of the
surface can be extracted quite clearly. It is $d_c\approx 2.35\pm0.10$ if
we choose $\rho=0.40$ in contrast to $d_c\to 2.0$ for uncorrelated
numbers. We illustrate the change of the cluster by 
cutting the system along the $x-t-$plane. Fig.\ref{cl3} shows such cuts
for $y=H/2$ where $a)$ denotes random costs and $b)$ corresponds to
strongest correlations.

\section{Summary}
In the present paper we have considered the effect of isotropically
correlated bond weights on the scaling behavior of both undirected and
directed paths in two and three dimensions. We found that in $d=2$ the
algebraic correlation of the disorder are relevant for a decay slower
than $1/r$ (i.e.\ $\rho=0$) and the roughness exponent $\nu$ and the
energy fluctuation exponent $\omega$ increase monotonously for
$\rho\ge0$. The Flory estimate (\ref{flory}) for the roughness exponent is
fulfilled as a lower bound of our numerical estimates. A precise value
for $\rho_c$, above which correlations modify the universality class, is
hard to estimate numerically, but we observe that it is close to and
slightly smaller than zero in $d=2$ and $d=3$, in agreement with the
Flory argument (\ref{flory}).

Moreover, the results in $d=2$ indicate that the scaling exponents are
independent from the directedness even in case of very strong
correlations. In contrast to this, in $d=3$ directed and undirected
lattices yield different results for $\rho>\rho_c\approx0$ indicating
that both cases constitute different universality classes. In $2d$ the
scaling relation $2\nu-\omega$ appears to remain valid even for stronge
correlations (although Galilean invariance is broken), whereas for $d=3$
we observe in the directed case significant deviations from it (in the
undirected case it appears to remain valid).

Finally we could exclude the possibility that optimal paths become
fractal for strong disorder correlations as long as they decay
algebraically. As a consequence $\nu$ stays smaller than one and
overhangs turn out to be irrelevant. This behavior changes if we
consider the following disorder correlations: $\langle(\eta_{\bf
j}-\eta_{{\bf j}+{\bf r}})^2\rangle\propto r^\alpha$, with $\alpha>0$,
which increase with distance $r=|{\bf r}|$. These type of correlators
are relevant for an effective model of dislocations in a vortex line
lattice \cite{vinokur}. In this case we found that the optimal paths are
fractals with a fractal dimension significantly larger than one and
depending on the value of $\alpha$ \cite{ref28}.

\acknowledgement
This work has been supported by the {\it Deutsche
Forschungsgemeinschaft} (DFG).

\appendix
\section[*]{Appendix}
\label{aclat} 

\begin{figure}[hbt]
\centerline{\epsfig{file=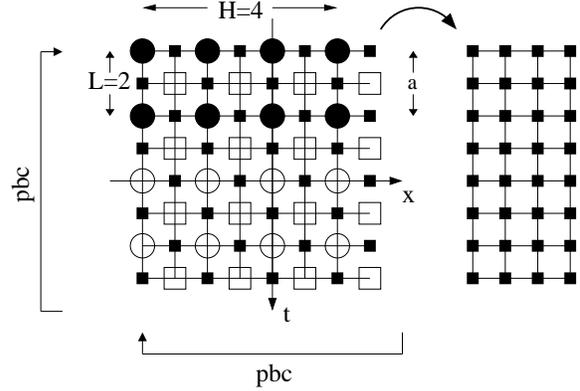,scale=0.72}}
\caption{\label{clat} First we stretch the lattice of size $(H\times 2L)$ 
resulting in $(2H\times 4L)$ and finally compress it again. Here 
{\Large$\bullet$} denotes sites that refer to nodes on a lattice of size
 $L$ and $H$, {\Large$\circ$} corresponds to virtual nodes which have to be
 introduced in order to follow the definition of the correlator
 (periodicity in each direction) and
 $\blacksquare$ refers to bonds. Squares denote positions that
 additionally arise by switching to a lattice spacing $a/2$. In 
 contrast to black squares the entries of white squares do not play any
 role concerning the generation of correlated random numbers.}
\end{figure}

Here we explain how to initialize the correlation function 
$g_\rho({\bf j})$ and we show that $N=2^dLH^{d-1}$ correlated
random numbers have to be created for our purposes. We need to
generate $2N$ uncorrelated random numbers and have to perform a Fourier
transformation twice. If we consider the simple square lattice of nodes
with lattice spacing $a$ we can immediately see that we have to
distinguish between positions of nodes and positions of bonds. The
latter build a simple lattice that is rotated towards the node lattice
by $\pi/4$ with a lattice spacing enlarged by a factor $\sqrt{2}$
(Fig.\ref{clat}). But we have to create correlations according to
positions of \emph{bonds}. Let us assume now this lattice to be a two
dimensional array. The most comfortable solution of the generation of
correlations due to bond positions is to keep the original lattice
structure, meaning we switch to a spacing $a/2$ and finally focus on
those positions within this array that correspond to bonds in our
lattice. Then we define the center of that lattice (array) and
initialize each position of the array by  $|{\bf j}|^{2\rho-1}$, where
$-H/2\le{\bf j}_i<H/2$ or  $-L/2\le{\bf j}_i<L/2$ is the position
vector. This array is called the correlator $g_\rho({\bf j})$. There are
two problems: by doing so, we generate twice the quantum of correlated
random numbers we really need (the information of the black squared
positions) and we still have to discuss how to define $g_\rho(0)$. As
mentioned in section II, we already have to take into account a
factor $2$ from the longitudinal expansion so that $N$ becomes
$N=2\cdot2^dLH^{d-1}$, where again all sites in the left part of Fig.\ref{clat}
would be occupied. In order to restrict memory usage we compress
the lattice along one axes (here the x-axes), whereby
we loose the information of all the positions indicated by open symbols.
By doing so, we avoid the necessity of defining $g_\rho(0)$ corresponding
to a virtual node at the origin and, consequently, being not part of the
right side of Fig.\ref{clat}. More important, $N$ is reduced by a factor 2. 
Note that the Fourier transform on the compressed lattice also yields 
the desired correlations.

\end{document}